\documentclass[12pt,twoside]{article}
\usepackage{fleqn,espcrc1}
\usepackage{axodraw}

\usepackage{graphicx}
\usepackage[figuresright]{rotating}

\newcommand{\be}{\begin{equation}}
\newcommand{\ee}{\end{equation}}
\newcommand{\bea}{\begin{eqnarray}}
\newcommand{\eea}{\end{eqnarray}}
\newcommand{\baa}{\begin{eqnarray*}}
\newcommand{\eaa}{\end{eqnarray*}}
\newcommand{\bary}{\begin{array}}
\newcommand{\eary}{\end{array}}
\newcommand{\bit}{\begin{itemize}}
\newcommand{\eit}{\end{itemize}}

\renewcommand{\thefootnote}{\fnsymbol{footnote}}%
\title{{\bf Trilinear gauge boson couplings and bilepton production
in the $\mbox{SU(3)}_C \otimes \mbox{SU(3)}_L
\otimes \mbox{U(1)}_N$ models}}
\author{Hoang Ngoc Long\address{{\it
 Laboratoire de Physique Th$\acute e$orique 
LAPTH, Chemin de Bellevue, B P 110, F-74941
Annecy-le-Vieux Cedex, France\\
{\rm and}\\
Institute of Physics, NCST, P. O. Box 429, Bo Ho, Hanoi 10000,
Vietnam}}
\footnote{E-mail address: hnlong@iop.ncst.ac.vn}
and
Dang Van Soa\address{{\it The Abdus Salam International Centre
for Theoretical Physics, Trieste 34100, Italy}}
\footnote{On leave from Department of Physics,
Hanoi University of Mining and Geology, Hanoi, Vietnam }}

\begin{document}

\maketitle

\vspace*{1cm}

\begin{abstract}
 The trilinear gauge boson couplings in
the $\mbox{SU(3)}_C \otimes \mbox{SU(3)}_L \otimes
\mbox{U(1)}_N$ (3 - 3 - 1)models are presented.
We find that new $Z_2$ does not interact with the
usual (in the standard model) gauge bosons $Z, W^\pm$.
Based on these results, production of new heavy
gauge bosons at high energy colliders such as
$e^+\ e^-$ is calculated. We show that the  
cross sections obtained in the 3 - 3 - 1 model with 
right-handed neutrinos can be
 {\it one order}  bigger than the same in the minimal
3 - 3 - 1 model.   
\end{abstract}
\vspace*{0.5cm}

PACS number(s):12.10.-g, 12.60.-i, 13.10.+q, 14.80.-j.\\

\newpage

\section{Introduction}
Although the standard model (SM)~\cite{gaw}
of electroweak interactions  has been verified to great
precision in the recent years at LEP, SLC
and other places, there remain a few unanswered
questions  concerning mainly  the mass spectrum and the 
generation structure of quarks and leptons.
In particular the question of the number of generations
remains open and few progress has been made towards
the understanding of the interrelation
between  generations. 
Amongst the possible models beyond the standard
one, from modest extensions to
GUTs, few address this question, the generations are usually
assumed to be a  replicate of the first one. 
The models based on the $\mbox{SU(3)}_C \otimes \mbox{SU(3)}_L
\otimes \mbox{U(1)}_N$ gauge
group~\cite{svs,pp,fr,flt,hnl}, are
interesting form this point of view. They
have the following intriguing features:
Firstly the models are anomaly free only if the number of
generations $ N$ is a multiple of three.  If further one adds the
condition of QCD asymptotic freedom, which
is valid only if the number of generations  of quarks
is to be less than 5, it follows that the number of
generations is equal to 3. The second characteristics
of the 3 - 3 - 1 models is that 
one generation of quarks is treated differently
from two others. This could lead to a natural 
explanation for the unbalancingly heavy top
quark. The possibility of the third generation being
different from the first two is not excluded  experimentally.
While the anomalous  behaviour of the parameters
$R_b$ and $A_b$ in the LEP data~\cite{da}
has more or less disappeared, 
the effects are now  1.8 $ \sigma$ away from
the SM value for $R_b$ and  3 $\sigma$ for $A_b$,
there is still room for  generation universality
breaking in the third generation. 
The third interesting feature is that the Peccei-Quinn
symmetry naturally occurs in these models~\cite{pal}. 
Finally, from a phenomenological point of view, 
the 3 - 3 - 1 models  are very interesting, they predict new physics
at a scale only slightly above the SM scale
 (typically TeV)
and even give upper bounds on the mass of some new particles.
Therefore the models
can be confirmed or ruled out in the
next generation of collider experiments from
Tevatron,  LHC, or a future linear collider,
 in  stark contrast with `` grand desert" scenarios in
Grand Unification Theories.

  Despite the extremely precise measurements of the SM
parameters, one important component
has not been tested directly with precision:
the non-abelian self-couplings of the weak gauge bosons.
The measurements performed at LEP1 have provided us
with an extremely accurate knowledge of the parameters
of the $Z$ gauge boson: its mass, partial widths, and
total width. There even is first evidence that the
contributions of gauge-boson loops to the gauge-boson
self-energies are indeed required~\cite{pzs}. Thus, an indirect
confirmation of the existence of the trilinear gauge boson
couplings (TGC's) has been obtained. 
With the excellent performance
of the LEP machine at high energy in last couple of years,
electroweak physics at LEP2 now trully merits the epithet
``precise''. The core measurements of the LEP2 program, the $W$
mass, and the vector boson self-couplings have been made with
precision better, in some cases substantially so, than
elsewhere. The mixing in the neutral gauge boson sector and the 
angular distributions as well as the $W$ helicities in the final
states of $W^+ W^-$ production have been searched for at 
LEP2~\cite{lep2}.
Deviation of non-abelian
couplings from expectation would signal new physics.
In addition, precise measurements of gauge boson self-interactions 
will provide important information on the nature
of electroweak symmetry breaking. The TGC's have been investigated
by many authors~\cite{gg,hpz,fb,rev}, and some direct tests
of these couplings have been made in~\cite{bel}. TGC's in the
beyond - the standard models, in which there exist  heavy
particles with mass much larger $m_W$ have been investigated
in~\cite{tai}. In the 3 - 3 - 1 models, the TGCs have the structure of the
standard model couplings, up to a coupling constant. 
Recent investigations have indicated that
signals of new gauge bosons in the models may be observed
at  the  CERN  LHC~\cite{london} or  Next Linear Collider 
(NLC)~\cite{ras}. 

  Our aim in this paper is to present TGC's in the
3 -- 3 -- 1 models and  use these couplings
to discuss new processes that could be measured
at future high energy
colliders. This paper is organized as follows.
In Sec. II we give a brief review of two models:
relation among {\it real} physical bosons and
gauge fields, which is necessary in getting of TGC's.
TGC's are given in this section. Sec.
III is devoted to bilepton production at high energy
collider $e^+ \ e^-$ and discussions are given in
the last section - section IV.
\section{A review of the 3 -- 3 -- 1 models }

 To frame the context, it is appropriate to briefly recall
some relevant features of two types of 3 -- 3 -- 1 models: 
the minimal model proposed by Frampton, Pisano and 
Pleitez~\cite{pp,fr}, and the model with right-handed (RH)
neutrinos~\cite{flt,hnl}.
\subsection{The minimal 3 -- 3 -- 1 model}

The model treats the leptons as  $\mbox{SU(3)}_L$
antitriplets~\cite{fr,dng}
\be
f^{a}_L = \left( \begin{array}{c}
               e^a_L\\ -\nu^a_L\\  (e^c)^a
               \end{array}  \right) \sim (1, \bar{3}, 0),
\label{l}
\ee
where $ a = 1, 2, 3$ is the generation index.

  Two of the three quark generations transform as triplets and
the third generation is treated differently - it belongs 
to an antitriplet:
\be
Q_{iL} = \left( \begin{array}{c}
                u_{iL}\\d_{iL}\\ D_{iL}\\
                \end{array}  \right) \sim (3, 3, -\frac{1}{3}),
\label{q}
\ee
\[ u_{iR}\sim (3, 1, 2/3), d_{iR}\sim (3, 1, -1/3),
D_{iR}\sim (3, 1, -1/3),\ i=1,2,\]
\be
 Q_{3L} = \left( \begin{array}{c}
                 d_{3L}\\ - u_{3L}\\ T_{L}
                \end{array}  \right) \sim (3, \bar{3}, 2/3),
\ee
\[ u_{3R}\sim (3, 1, 2/3), d_{3R}\sim (3, 1, -1/3), T_{R}
\sim (3, 1, 2/3).\]
Of the nine gauge bosons  $W^a (a = 1, 2, ..., 8)$ and
$B$ of $SU(3)_L$ and $U(1)_N$, four
are light: the photon ($A$), $Z$ and $W^\pm$. The remaining
five correspond to new heavy gauge bosons 
$Z_2,\ Y^\pm$ and the doubly charged
bileptons $X^{\pm \pm}$. They are expressed  in terms of
$W^a$ and $B$  as~\cite{dng}:
\renewcommand{\thefootnote}{\fnsymbol{footnote}}%
\footnote{The leptons may be assigned to  a
triplet as in~\cite{pp}, however the two models
are mathematically identical.}
\bea
\sqrt{2}\ W^+_\mu &=& W^1_\mu - iW^2_\mu ,
\sqrt{2}\ Y^+_\mu = W^6_\mu - iW^7_\mu ,\nonumber\\
\sqrt{2}\ X_\mu^{++} &=& W^4_\mu - iW^5_\mu,
\label{idmin1}
\eea
\bea
A_\mu  &=& s_W  W_{\mu}^3 + c_W\left(\sqrt{3}\ t_W\ W^8_{\mu}
+\sqrt{1- 3\ t^2_W}\  B_{\mu}\right),\nonumber\\
Z_\mu  &=& c_W  W_{\mu}^3 - s_W\left(\sqrt{3}\ t_W\ W^8_{\mu}
+\sqrt{1- 3\ t^2_W}\  B_{\mu}\right),\nonumber\\
Z'_\mu &=&-\sqrt{1- 3\ t^2_W}\ \ W^8_{\mu}
+ \sqrt{3}\ t_W\ B_{\mu},
\label{apstat}
\eea
where we use the following notations:
$s_W \equiv \sin \theta_W $ and $t_W \equiv \tan \theta_W$.
The {\it physical} states are a mixture of $Z$ and $Z'$:
\bea
Z_1  &=&Z\cos\phi - Z'\sin\phi,\nonumber\\
Z_2  &=&Z\sin\phi + Z'\cos\phi,\nonumber
\eea
where $\phi$ is a mixing angle.

  Symmetry breaking and fermion mass generation can be
achieved by three scalar  $\mbox{SU(3)}_L$ triplets and a sextet
\bea
&\mbox{SU}(3)_{C}&\hspace*{-0.2cm}\otimes \ \mbox{SU}(3)_{L}\otimes
\mbox{U}(1)_{N}\nonumber \\
&\downarrow      &\hspace*{-0.8cm}\langle \Phi \rangle   \nonumber \\
&\mbox{SU}(3)_{C}&\hspace*{-0.2cm}\otimes \ \mbox{SU}(2)_{L}\otimes
\mbox{U}(1)_{Y}\nonumber \\
&\downarrow      &\hspace*{-0.8cm}\langle \Delta \rangle, \langle
\Delta' \rangle, \langle \eta \rangle\nonumber   \\
&\mbox{SU}(3)_{C}&\hspace*{-0.2cm}\otimes \ \mbox{U}(1)_{Q},
\nonumber
\label{ssb2}
\eea
where the minimally required scalar multiplets
are summarized as
\bea
\Phi & = &\left( \begin{array}{c}
                \phi^{++}\\ \phi^+\\ \phi^{,o}\\
                \end{array}  \right) \sim (1, 3, 1),\nonumber \\
\label{mh1}
\Delta & =& \left( \begin{array}{c}
                \Delta^+_1\\ \Delta^o\\ \Delta^-_2\\
                \end{array}  \right) \sim (1, 3, 0),\nonumber \\
\Delta' & =& \left( \begin{array}{c}
                \Delta^{'o}\\ \Delta^{'-}\\ \Delta^{'--}\\
                \end{array}  \right) \sim (1, 3, -1),\nonumber \\
\eta & = & \left( \begin{array}{ccc}
\eta^{++}_1 & \eta^+_1/ \sqrt{2} & \eta^o/ \sqrt{2}\\
\eta^+_1/ \sqrt{2} & \eta^{'o} & \eta^-_2/ \sqrt{2}\\
\eta^o/ \sqrt{2} & \eta^-_2/ \sqrt{2} & \eta^{--}_2
                \end{array}  \right) \sim (1, 6, 0).\nonumber
\eea
The vacuum expectation value (VEV) $\langle \Phi^T \rangle
= ( 0, 0, u/ \sqrt{2})$ yields masses for the
exotic quarks, the heavy neutral gauge boson ($Z_2$) and
two charged gauge bosons ($X^{++}, Y^+$). The
masses of the standard gauge bosons and the ordinary fermions
are related to the VEVs of the other scalar fields,
$\langle \Delta^o \rangle = v/ \sqrt{2},
\langle \Delta'^o  \rangle = v'/ \sqrt{2}$ and
$\langle \eta^o \rangle = \omega/ \sqrt{2},\
\langle \eta^{'o} \rangle =0 $. 
 In order to be consistent with the low energy phenomenology
we have to assume that $u \gg\ v,\ v',\ \omega$.
The masses of gauge bosons are explicitly:
\be
m^2_W=\frac{1}{4}g^2(v^2+v^{'2}+\omega^2),\
M^2_Y=\frac{1}{4}g^2(u^2+v^2+\omega^2),
M^2_X=\frac{1}{4}g^2(u^2+v^{'2}+4 \omega^2),
\label{mnhb}
\ee
and
\bea
m_{Z}^2   &=&\frac{g^2}{4 c_W^2}(v^2+v^{'2}+\omega^2)=
\frac{m_W^2}{c_W^2},\nonumber \\
M_{Z'}^2 &=&\frac{g^2}{3}\left[\frac{c^2_W}{1
- 4 s^2_W} u^2 + \frac{1 - 4 s^2_W}{4
c^2_W}( v^2 + v^{'o} + \omega^2 )
+ \frac{3 s^2_W}{1
- 4 s^2_W} v^{'2}\right].
\label{masmat}
\eea

  Expressions in (\ref{mnhb}) yields a splitting 
on the bileptons masses~\cite{lng}
\be
| M_X^2 - M_Y^2 | \leq 3\  m_W^2.
\label{maship}
\ee

  By matching the gauge coupling constants we get a relation
between $g$ and $g_N$ -- the couplings associated with
$\mbox{SU(3)}_L$ and $\mbox{U(1)}_N$, respectively:
\be
\frac{g_N^2}{g^2} = \frac{6 \ s^2_W(M_{Z_2})}{1 -
4  s^2_W(M_{Z_2})},
\label{coupm}
\ee
where $e = g\  s_W$ is the same as in the SM.

  Combining constraints from direct searches
and neutral currents, one obtains a range
for the mixing angle~\cite{dng} $- 1.6
\times 10^{-2} \le \phi \le 7 \times 10^{-4}$ and 
a lower bound on $M_{Z_2}\ge 1.3 $ TeV.
Such a small mixing angle can safely be neglected.
Adding the constraints from ``wrong'' muon  decay experiments
one also obtains a range for the
new gauge charged bosons:  $ M_{Y^+} \ge 230 $~ GeV.
By computing the oblique parameters $S$ and $T$,
a lower bound of 367 GeV for the mass of the singly
charged bilepton $Y^+$ is derived~\cite{fraha}.
Combining this with the mass splitting given in (\ref{maship})
we obtain a lower bound around 400 GeV for the mass of the
doubly charged bilepton ($X^{++})$. However the most
stringent limit  on the mass of doubly charged bilepton
is derived from constraints on fermion pair production
at LEP and lepton-flavour violating charged lepton 
decay~\cite{tuj}:$M_{X^{++}} > 740$ GeV.
With  the new atomic parity violation
in cesium, one gets a lower
bound for the $Z_2$ mass~\cite{ltrun}: $M_{Z_2} > 1.2 $ TeV.
From symmetry breaking it follows that the masses of the new
charged gauge bosons $Y^\pm, \ X^{\pm\pm}$ are less than a
half of $M_{Z_2}$,
the allowed decay $Z_2 \rightarrow
X^{++}\ X^{--}$ with $X^{\pm\pm} \rightarrow
2 l^\pm$ provides a unique signature in future colliders.

  The TGC's in this model are obtained from the part of the Lagrangian
describing the self-interactions of gauge fields.
\be
{\cal L}_{TGC} = - g\ f_{abc}\ \partial_\mu W_\nu^a \
W^{b \mu}\ W^{c \nu}, \ a, b, c = 1, 2, ..., 8.
\ee
Expressing $W^a \ (a = 1, 2, ..., 8)$ in terms of physical
fields using Eqs (\ref{idmin1}) and (\ref{apstat}), a 
straightforward but cumbersome calculation leads to
\bea
\frac{i}{g}{\cal L}_{TGC}^{min} &=& s_W \left[ A^\nu (
W^-_{\mu \nu} W^{+\mu} - W^+_{\mu \nu} W^{-\mu} ) +
A_{\mu \nu} W^{-\mu} W^{+\nu}\right]  \nonumber \\
 & & + c_W \left[ Z^\nu (
W^-_{\mu \nu} W^{+\mu} - W^+_{\mu \nu} W^{-\mu} ) +
Z_{\mu \nu} W^{-\mu} W^{+\nu}\right]  \nonumber \\
 & & + s_W \left[ A^\nu (
Y^-_{\mu \nu} Y^{+\mu} - Y^+_{\mu \nu} Y^{-\mu} ) +
A_{\mu \nu} Y^{-\mu} Y^{+\nu}\right] \nonumber \\
 & & - \frac{(c_W + 3 s_W t_W)}{2} \left[ Z^\nu (
Y^-_{\mu \nu} Y^{+\mu} - Y^+_{\mu \nu} Y^{-\mu} ) +
Z_{\mu \nu} Y^{-\mu} Y^{+\nu}\right] \nonumber \\
 & & + 2 s_W \left[ A^\nu (
X^{--}_{\mu \nu} X^{++\mu} - X^{++}_{\mu \nu} X^{--\mu} ) +
A_{\mu \nu} X^{--\mu} X^{++\nu}\right] \nonumber \\
 & & + \frac{(c_W - 3 s_W t_W)}{2} \left[ Z^\nu (
X^{--}_{\mu \nu} X^{++\mu} - X^{++}_{\mu \nu} X^{--\mu} ) +
Z_{\mu \nu} X^{--\mu} X^{++\nu}\right] \nonumber \\
& & - \frac{\sqrt{3(1 - 3 t_W^2)}}{2} \left[ Z'^\nu (
Y^-_{\mu \nu} Y^{+\mu} - Y^+_{\mu \nu} Y^{-\mu} ) +
Z'_{\mu \nu} Y^{-\mu} Y^{+\nu}\right] \nonumber \\
 & &  - \frac{\sqrt{3(1 - 3 t_W^2)}}{2} \left[ Z'^\nu (
X^{--}_{\mu \nu} X^{++\mu} - X^{++}_{\mu \nu} X^{--\mu} ) +
Z'_{\mu \nu} X^{--\mu} X^{++\nu}\right] \nonumber \\
 & & + \frac{1}{\sqrt{2}} \left[ X^{--\nu} (
Y^+_{\mu \nu} W^{+\mu} - W^+_{\mu \nu} Y^{+\mu} ) +
X^{--}_{\mu \nu} Y^{+\mu} W^{+\nu}\right] \nonumber \\
 & & + \frac{1}{\sqrt{2}} \left[ X^{++\nu} (
W^-_{\mu \nu} Y^{-\mu} - Y^-_{\mu \nu} W^{-\mu} ) +
X^{++}_{\mu \nu} W^{-\mu} Y^{-\nu}\right],
\label{lt}
\eea
where $W_{\mu \ \nu} \equiv  \partial_\mu W_\nu -
\partial_\nu  W_\mu$.
The coupling constants for all trilinear couplings are
summarized in  Table 1.
\begin{table}[htb]\caption{Trilinear couplings in the minimal
3 -- 3 -- 1 model.}
\begin{tabular}{|c|c|}  \hline \hline
Vertex & coupling constant/e   \\  \hline \hline
$\gamma W^+ W^-$  & 1\\  \hline
$Z W^+ W^-$ &$1/t_W$ \\ \hline
$\gamma Y^+ Y^-$&$1$\\ \hline
$Z Y^+ Y^-$ & $- (1 + 2 s_W^2)/\sin 2\theta_W$\\  \hline
$\gamma X^{++} X^{--}$&$ 2 $\\ \hline
$Z X^{++} X^{--}$&$(1 - 4 s_W^2)/\sin 2\theta_W$\\ \hline
$Z' Y^+ Y^-$&$ - \sqrt{3(1 - 4 s_W^2)}/\sin
2\theta_W$\\ \hline
$Z' X^{++} X^{--}$&$ -  \sqrt{3(1 - 4 s_W^2)}/\sin
2\theta_W$\\ \hline
$X^{--} Y^+ W^+$&$ 1/(\sqrt{2}\  s_W)$\\ \hline
$X^{++} W^- Y^-$&$ 1/(\sqrt{2} \ s_W)$\\ \hline
\end{tabular}\\[2pt]
\end{table}

As we can see from Table 1, the $Z'$ does not interact with
the usual gauge bosons: photon, $W^\pm$ and $Z$.
Strictly speaking, the  new neutral gauge boson $Z_2$
interacts very weakly with the usual SM bosons since 
the mentioned coupling constants are proportional
to $\sin \phi$.
The SM trilinear gauge boson couplings are recovered in (\ref{lt}).
\subsection{The model with RH neutrinos}
 In this model the leptons are in triplets, and the third member 
is a RH neutrino~\cite{flt,hnl}:
\be
f^{a}_L = \left( \begin{array}{c}
               \nu^a_L\\ e^a_L\\ (\nu^c_L)^a
\end{array}  \right) \sim (1, 3, -1/3), e^a_R\sim (1,
1, -1).
\label{l2}
\ee

The first two generations of quarks are in antitriplets while the
third one is in a triplet:
\be
Q_{iL} = \left( \begin{array}{c}
                d_{iL}\\-u_{iL}\\ D_{iL}\\
                \end{array}  \right) \sim (3, \bar{3}, 0),
\label{q}
\ee
\[ u_{iR}\sim (3, 1, 2/3), d_{iR}\sim (3, 1, -1/3),
D_{iR}\sim (3, 1, -1/3),\ i=1,2,\]
\be
 Q_{3L} = \left( \begin{array}{c}
                 u_{3L}\\ d_{3L}\\ T_{L}
                \end{array}  \right) \sim (3, 3, 1/3),
\ee
\[ u_{3R}\sim (3, 1, 2/3), d_{3R}\sim (3, 1, -1/3), T_{R}
\sim (3, 1, 2/3).\]
The doubly charged bileptons of the minimal model are replaced
here by complex neutral  ones:
\bea
\sqrt{2}\ W^+_\mu &=& W^1_\mu - iW^2_\mu ,
\sqrt{2}\ Y^-_\mu = W^6_\mu - iW^7_\mu ,\nonumber\\
\sqrt{2}\ X_\mu^o &=& W^4_\mu - iW^5_\mu.
\eea
For a  shorthand notation, hereafter we
will use $X^o \equiv X$.

  The {\it physical} neutral gauge bosons are again related
to $Z, Z'$
through the mixing angle $\phi$. Together with the
photon, these are defined as follows~\cite{hnl}:
\begin{eqnarray}
A_\mu  &=& s_W  W_{\mu}^3 + c_W\left(-
\frac{t_W}{\sqrt{3}}\ W^8_{\mu}
+\sqrt{1-\frac{t^2_W}{3}}\  B_{\mu}\right),
\nonumber\\
Z_\mu  &=&  c_W  W^3_{\mu} - s_W\left(
-\frac{t_W}{\sqrt{3}}\ W^8_{\mu}+
\sqrt{1-\frac{t_W^2}{3}}\  B_{\mu}\right),  \\
Z'_\mu &=& \sqrt{1-\frac{t_W^2}{3}}\
W^8_{\mu}+\frac{t_W}{\sqrt{3}}\ B_{\mu}.\nonumber
\label{apstat1}
\end{eqnarray}
The symmetry breaking can be achieved with just three
$\mbox{SU}(3)_L$ triplets
\bea
&\mbox{SU}(3)_{C}&\hspace*{-0.2cm}\otimes \ \mbox{SU}(3)_{L}\otimes
\mbox{U}(1)_{N}\nonumber \\
&\downarrow      &\hspace*{-0.8cm}\langle \chi \rangle   \nonumber \\
&\mbox{SU}(3)_{C}&\hspace*{-0.2cm}\otimes \ \mbox{SU}(2)_{L}\otimes
\mbox{U}(1)_{Y}\nonumber \\
&\downarrow      &\hspace*{-0.8cm}\langle \rho \rangle, \langle
\eta \rangle   \\
&\mbox{SU}(3)_{C}&\hspace*{-0.2cm}\otimes \ \mbox{U}(1)_{Q},
\nonumber
\label{ssb2}
\eea
where
\bea
\chi & = &\left( \begin{array}{c}
                \chi^o\\ \chi^-\\ \chi^{,o}\\
                \end{array}  \right) \sim (1, 3, -1/3),\\
\label{h1}
\rho & =& \left( \begin{array}{c}
                \rho^+\\ \rho^o\\ \rho^{,+}\\
                \end{array}  \right) \sim (1, 3, 2/3),\\
\eta & =& \left( \begin{array}{c}
                \eta^o\\ \eta^-\\ \eta^{,o}\\
                \end{array}  \right) \sim (1, 3, -1/3).\\
\eea
The necessary VEVs are
\be
\langle\chi \rangle^T = (0, 0, \omega/\sqrt{2}),\
\langle\rho \rangle^T = (0, u/\sqrt{2}, 0),\
\langle\eta \rangle^T = (v/\sqrt{2}, 0, 0).
\label{vev}
\ee
Here the electric charge is defined:
\be
Q=\frac{1}{2}\lambda_3-\frac{1}{2\sqrt{3}}\lambda_8+N.
\label{charge}
\ee
The VEV $\langle \chi \rangle$ generates masses for the
exotic 2/3 and --1/3
quarks, while the  VEVs $\langle \rho \rangle$ and
$\langle \eta \rangle$ generate
masses for all ordinary leptons and quarks.
Neutrinos, however, are still massless. After symmetry
breaking the gauge bosons gain  masses
\be
m^2_W=\frac{1}{4}g^2(u^2+v^2),\
M^2_Y=\frac{1}{4}g^2(v^2+\omega^2),
M^2_X=\frac{1}{4}g^2(u^2+\omega^2),
\label{rhb}
\ee
and
\bea
m_{Z}^2   &=&\frac{g^2}{4 c_W^2}(u^2+v^2)=
\frac{m_W^2}{c_W^2},\nonumber \\
M_{Z'}^2 &=&\frac{g^2}{4(3-4s_W^2)}\left[4
\omega^2+ \frac{u^2}{c_W^2}
+ \frac{v^2(1-2s_W^2)^2}{c_W^2}\right].
\label{masmat}
\eea

  In order to be consistent with the low energy phenomenology
we have to assume that $\langle \chi \rangle \gg\
\langle \rho \rangle,\ \langle \eta \rangle$
such that $m_W \ll M_X, M_Y$.

  In this model the analog of formula (\ref{coupm}) for the 
ratio of coupling constants is :
\be
\frac{g_N^2}{g^2} = \frac{18 \ s^2_W(M_{Z_2})}{3 -
4 s^2_W(M_{Z_2})}.
\ee

  The symmetry-breaking hierarchy gives us splitting on the 
bileptons masses~\cite{til}
\be
| M_X^2 - M_Y^2 | \leq m_W^2.
\label{mashipr}
\ee
Therefore in the future studies
it is acceptable to put $M_X \simeq M_Y$.

  The constraint on the $Z - Z'$ mixing based on
the $Z$ decay, is given~\cite{hnl}:
$-2.8 \times 10^{-3} \le \phi \le 1.8 \times 10^{-4}$, and
in this model we have not  a limit for $\sin^2 \theta_W$.
From the data on parity violation in the cesium atom,
one gets a lower bound on the $Z_2$ mass at a range
from 1.4 TeV to 2.6 TeV~\cite{ltrun}.  The muon decay
data~\cite{pdg} gives a lower bound for the $Y$ boson mass: 
230 GeV (90 \% CL). Analyzing radiative correction based on 
the $S$ and $T$ parameters gives similar results~\cite{til}:
$M_{Y^+} \ge 230\ {\rm GeV}, M_{X} \ge 240$
GeV.

  Repeating the procedure for deriving
the trilinear interactions of gauge bosons, one obtains:
\bea
\frac{i}{g}{\cal L}_{TGC}^{rhn} &=& s_W \left[ A^\nu (
W^-_{\mu \nu} W^{+\mu} - W^+_{\mu \nu} W^{-\mu} ) +
A_{\mu \nu} W^{-\mu} W^{+\nu}\right]  \nonumber \\
 & & + c_W \left[ Z^\nu (
W^-_{\mu \nu} W^{+\mu} - W^+_{\mu \nu} W^{-\mu} ) +
Z_{\mu \nu} W^{-\mu} W^{+\nu}\right]  \nonumber \\
 & & + s_W \left[ A^\nu (
Y^-_{\mu \nu} Y^{+\mu} - Y^+_{\mu \nu} Y^{-\mu} ) +
A_{\mu \nu} Y^{-\mu} Y^{+\nu}\right] \nonumber \\
 & & + \frac{(c_W - s_W t_W)}{2} \left[ Z^\nu (
Y^-_{\mu \nu} Y^{+\mu} - Y^+_{\mu \nu} Y^{-\mu} ) +
Z_{\mu \nu} Y^{-\mu} Y^{+\nu}\right] \nonumber \\
 & & - \frac{(c_W + s_W t_W)}{2} \left[ Z^\nu (
X_{\mu\nu} X^{*\mu} - X^{*}_{\mu\nu} X^{\mu} ) +
Z_{\mu \nu} X^\mu X^{*\nu} \right] \nonumber \\
& & - \frac{\sqrt{3 - t_W^2}}{2} \left[ Z'^\nu (
Y^-_{\mu \nu} Y^{+\mu} - Y^+_{\mu \nu} Y^{-\mu} ) +
Z'_{\mu \nu} Y^{-\mu} Y^{+\nu}\right] \nonumber \\
 & &  - \frac{\sqrt{3 - t_W^2}}{2}  \left[ Z'^\nu (
X_{\mu \nu} X^{*\mu} - X^*_{\mu \nu} X^\mu ) +
Z'_{\mu \nu} X^\mu X^{*\nu}\right] \nonumber \\
 & & + \frac{1}{\sqrt{2}} \left[ X^{\nu} (
W^-_{\mu \nu} Y^{+\mu} - Y^+_{\mu \nu} W^{-\mu} ) +
X_{\mu \nu} W^{-\mu} Y^{+\nu}\right] \nonumber \\
 & & + \frac{1}{\sqrt{2}} \left[ X^{*\nu} (
Y^-_{\mu \nu} W^{+\mu} - W^+_{\mu \nu} Y^{-\mu} ) +
X^*_{\mu \nu} Y^{-\mu} W^{+\nu}\right].
\eea

The coupling constants for the TGC's
in this model are listed in Table 2
\begin{table}[htb]\caption{Trilinear couplings in the
3 -- 3 -- 1 model with RH neutrinos.}
\begin{tabular}{|c|c|}  \hline \hline
Vertex & coupling constant/e   \\  \hline \hline
$\gamma W^+ W^-$  & $1$\\  \hline
$Z W^+ W^-$ &$ 1/t_W$ \\ \hline
$\gamma Y^+ Y^-$&$1$\\ \hline
$Z Y^+ Y^-$ & $ 1 /  \tan 2\theta_W$\\  \hline
$Z X  X^*$&$- 1 /\sin 2\theta_W$\\ \hline
$Z' Y^+ Y^-$&$ -  \sqrt{3 - 4 s_W^2}/\sin
2\theta_W$\\ \hline
$Z' X X^*$&$ -  \sqrt{3 - 4 s_W^2}/\sin
2\theta_W$\\ \hline
$X  W^- Y^+ $&$ 1/(\sqrt{2}\  s_W)$\\ \hline
$X^* Y^- W^+$&$ 1/(\sqrt{2} \ s_W)$\\ \hline
\end{tabular}\\[2pt]
\end{table}

 As we can see again from Table 2, the $Z'$ does not interact with
the usual gauge bosons: photon, $W^\pm$ and $Z$. As expected
there is no coupling of the  photon with the neutral  bosons $X$,
and the SM couplings remain undefected. It must be noted
that the coupling of the $Z_2$ to the new gauge bosons
is much stronger than that in the minimal model 
which was supprerssed by a factor $( 1 - 4 s_W^2)$.

\section{Bilepton production  at high energy
colliders }
Recently production of doubly-charged
vector bileptons
in high energy collision has been widely
discussed both in generic models~\cite{rizzo,cd}
and in the minimal model~\cite{ras,fl}.
The production of bileptons in the 3 - 3 - 1 model
is particularly relevant for colliders in the TeV range
since the models predict new gauge bosons at the same
scale. Furthermore the present constraints on bilepton masses are
not very stringent~\cite{con}: 
the constraint 
is only $M_Y \ge 230$ GeV from the muon decay experiment.
One of the strongest limits on the bilepton mass comes
from the fermion pair production and lepton-flavour
violating charged lepton decays~\cite{tuj} at about
750 GeV. The current experimental lower limit on $M_{X^{--}}$ is
claimed to be 850 GeV~\cite{wil} (However, such  a lower 
limit can be derived from oblique corrections assuming
the Higgs mass to be equal to 300 GeV~\cite{fraha}). 
In this section, we present the cross sections for  the
production of  two bileptons in both types of 3 - 3 - 1 models. 
In the models we are considering one 
has to include, in addition to the
photon and $Z$ exchange, the  contribution from the new neutral
gauge boson $Z_2$  as well as from internal lepton  exchange
(in $t$ channel).
These were not considered in the general calculation
of~\cite{cd} and we will see that in many cases,
especially for the model with RH neutrinos,
the $Z_2$ has a significant contribution.  

  There are four modes for the bileptons discovery:
$e^+ e^-, e^- e^-, \mu^- \mu^-$ and $\gamma \gamma$.
The $e^- e^-$ and $\mu^- \mu^- $ running modes of the linear
colliders are particularly suitable for discovering of doubly
charged bileptons. However we have to wait for these modes. 
In this paper we concentrate on production
of the bileptons at  high energy $e^+ e^-$ colliders 
from 1 TeV (NLC) to 3 TeV, such as the
CLIC project designed at CERN.

\subsection{ Production  of bileptons in  the minimal model}

  Bileptons (singly and doubly charged) can be produced
in high energy $e^+ \ e^-$ colliders and 
this process has been examined (see report of Dawson in~\cite{da}).
In this model the intermediate states are photon, $Z$ and
{\it new} $Z_2$ bosons in the $s$ channel and
neutrino/electron in $t$ channel. It must be emphasized 
that in~\cite{cd} the intermediate states consist
only of the first two neutral gauge bosons, while in our case,
the $Z_2$ also gives a contribution.

\vspace*{0.9cm}

\begin{center}
\begin{picture}(260,50)(-5,0)
\ArrowLine(-10,50)(11,10)
\ArrowLine(11,10)(-10,-30)
\Photon(11,10)(80,10){2}{6}
\Photon(80,10)(100,50){2}{4}
\Photon(80,10)(100,-30){2}{4}
\Text(45,18)[]{$\gamma, Z, Z_2 $}
\Text(-16,54)[]{$e^-$}
\Text(-16,-34)[]{$e^+$}
\Text(112,54)[]{$Y^-$}
\Text(112,-34)[]{$Y^+$}
\ArrowLine(150,50)(195,40)
\ArrowLine(195,-20)(150,-30)
\ArrowLine(195,-20)(195,40)
\Photon(195,40)(235,50){2}{4}
\Photon(195,-20)(235,-30){2}{4}
\Text(145,54)[]{$e^- $}
\Text(185,10)[]{$\nu_e$}
\Text(145,-34)[]{$e^+$}
\Text(245,54)[]{$Y^-$}
\Text(245,-34)[]{$Y^+$}
\Text(90,-80)[]{ Figure 1: Feynman diagram for 
$\ e^+ \ e^- \rightarrow  Y^+ \ Y^-  $}
\Text(90,-90)[]{ in the  3 - 3 - 1 models }
\end{picture}
\end{center}
\vspace*{4cm}

First we consider the production of singly charged bileptons:
\be
e^-(k,\lambda) \ + \ e^+(k',\lambda ') \rightarrow 
Y^-(p,\tau) \ + \ Y^+(p',\tau '),
\label{pr1}
\ee
where the first and the second letters in parentheses stand for 
the momentum and the helicity of the particle, respectively.
The Feynman diagram for the full process is depicted in Fig. 1.

  The amplitude for this process due to neutrino,
$\gamma, Z$ and $Z_2$ is given (in the notation of~\cite{mou})
\be
{\cal M}_{fi} = {\cal M}^\nu_{fi} +  {\cal M}^\gamma_{fi}
+  {\cal M}^Z_{fi}  + {\cal M}^{Z_2}_{fi}.
\ee
The neutrino exchange part is written as
\begin{equation}
{\cal M}^\nu_{fi} = -\frac{e^2}{4 t s_W^2}\bar{v}({\bf k}')
\epsilon'\!\!\!/(k\!\!\!/ - p\!\!\!/)\epsilon\!\!\!/
( 1 - \gamma^5) u({\bf k}).
\end{equation}

The diagrams with $\gamma, Z, Z_2$ intermediate lines involve
the three-boson vertices defined in Subsect. {\bf 2.1} and
 the new $Z_2 e^+ e^-$ (vector and axial) vertices
\begin{equation}
g^V_{Z_2 ee}  =  \frac{\sqrt{3}}{2} \sqrt{1 - 4 s_W^2},\
g^A_{Z_2 ee}  =  - \frac{1}{2\sqrt{3}} \sqrt{1 - 4 s_W^2}.
\end{equation}

\setcounter{figure}{1}
\begin{figure*}[t]
\begin{center}
\includegraphics[width=10cm,height=8cm]{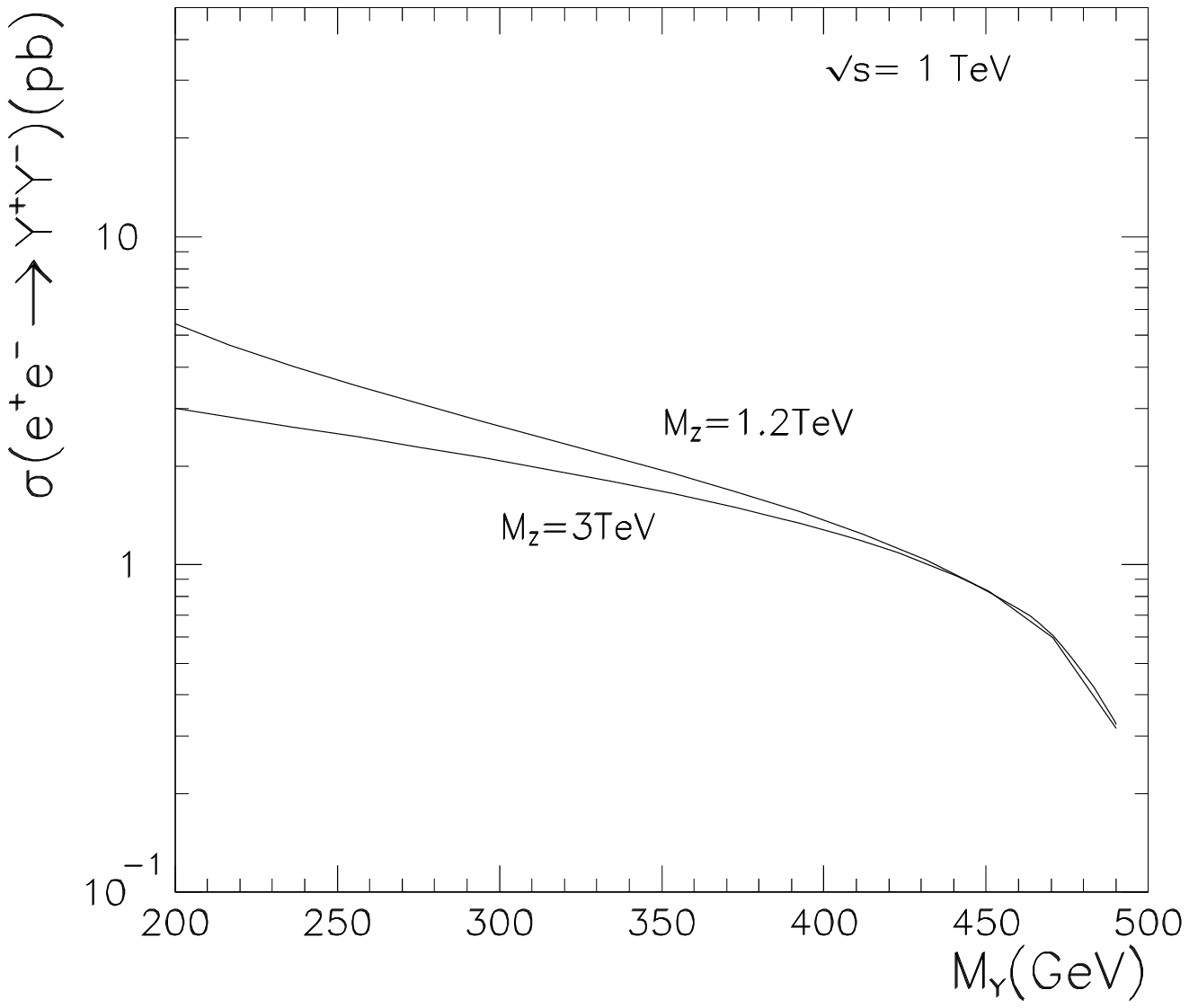}
\caption{\label{lim}{\em Cross section $\sigma (e^+ e^- \rightarrow
Y^+ Y^-$) in the minimal 3 - 3 - 1 model as a function of $M_Y$.}}
\end{center}
\end{figure*}

Hence contributions of the diagrams with gauge boson exchange
 are given by
\bea
{\cal M}^\gamma_{fi} &=& -\frac{e^2}{s} Q_Y\bar{v}({\bf k}')
\gamma^\mu u({\bf k}) {\cal V}^\gamma_\mu,
\label{ap1}\\
{\cal M}^Z_{fi} &=& \frac{e^2}{s-m_Z^2}\bar{v}({\bf k}')
\gamma^\mu ( a -  b\gamma^5) u({\bf k}) {\cal V}^Z_\mu,
\label{ap2}
\\
{\cal M}^{Z_2}_{fi} &=& \frac{e^2}{s - M_{Z_2}^2}\bar{v}({\bf k}')
\gamma^\mu ( a' -  b'\gamma^5) u({\bf k}) {\cal V}^{Z_2}_\mu,
\label{ap3}
\eea
 where
\bea
a&=& \frac{ ( 1 - 4 s_W^2)( 1 + 2 s_W^2) }{ 4
c_W^2 \sin 2\theta_W},\
b = - \frac{ ( 1 + 2 s_W^2) }{ 4 c_W^2 \sin 2\theta_W},
\label{ab1}
\\
a'&=&  \frac{3(- 1 + 4 s_W^2) }{ 4 c_W^2 \sin 2\theta_W},\
b'= \frac{ (- 1 + 4 s_W^2) }{ 4 c_W^2 \sin 2\theta_W}.
\label{abpr1}
\eea
With the notations of Ref~\cite{mou} (see fig. 1 there),
${\cal V}^V_\mu$ is defined as follows
\[{\cal V}^V_\mu = g^V [\epsilon . \epsilon'
(p - p')_\mu - 2 \epsilon'.p \epsilon_\mu +
2 \epsilon.p' \epsilon'_\mu]. \]
As usual, we have used $s = (k + k')^2 = (p + p')^2,
 t =  (k - p)^2 = (p' - k')^2$.

In the high energy-limit, $s \gg m_Z^2,
M^2_{Z_2}$ unitarity considerations of partial wave amplitudes
imply cancellations among the various diagrams to control
the bad  high-energy behaviour of each amplitude.
The sum of the amplitudes for the production of longitudinal
gauge bosons will tend to zero. For the production
of singly charged bilepton (\ref{pr1}), the contributions from
$\nu$ and $\gamma$ exchange are the same as in
$e^+ \ e^- \rightarrow W^+\  W^-$ in the SM.
Therefore contributions from $Z$ and $Z_2$ exchanges
should be equal to $Z$ exchange in the SM.
From (\ref{ab1}) and (\ref{abpr1}) it is easy to check that 
such is the case, indeed in this model we have:
\begin{eqnarray}
a + a' &=& \frac{-1 + 4 s_W^2}{4 s_W c_W} = a_{SM}\\
b + b' &=& -\frac{1}{4 s_W c_W} = b_{SM}.
\end{eqnarray}

 It is convenient to decompose the amplitude in  the
helicity basis: the helicity of the electron (positron)
is denoted by $\lambda = \pm \frac 1 2 (\lambda' =
- \lambda)$ while the helicities of the $Y^-$ and
$Y^+$ by $\tau = \pm 1, 0$ and $\tau' = \pm 1, 0$),
respectively. 
They are given in Table 3,
where the first row corresponds to the lepton exchange diagram
and the second row to the gauge bosons exchange diagrams:
\begin{table}[htb]\caption{ The helicity amplitudes for
$e^+ \ e^- \rightarrow Y^+ Y^-$}
\begin{tabular}{|c|c|c|c|c|}  \hline
 &$\tau = \tau' = \pm 1$& $\tau = - \tau' = \pm 1$&
$\tau = \tau' = 0$&$ \tau = 0, \tau' = \pm 1, \epsilon = 1$
 \\
 &$ -\frac{1}{2} e^2 s \lambda \sin\vartheta$&
$ -\frac{1}{2} e^2 s \lambda \sin\vartheta$&
$ -\frac{1}{2} e^2 s \lambda \sin\vartheta$&
$-\frac{e^2s\lambda}{2\sqrt{2}}(\tau'\cos\vartheta-2\lambda)$
 \\  \hline \hline
$\frac{2 \lambda -1}{4 ts_W^2}$&$ \cos\vartheta -\beta_Y$&
$-\cos\vartheta - 2 \tau \lambda$&$\frac{s}{2M_Y^2}
[\cos\vartheta $&
$\frac{\sqrt{s}}{2M_Y}[\cos\vartheta (1+\beta_Y^2) 
$\\  
&    &   &$- \beta_Y (1+\frac{2M_Y^2}{s})]$
&$-2\beta_Y] -\frac{2M_Y}{\sqrt{s}}\frac{\tau'\sin^2\vartheta}{(
\tau'\cos\vartheta -2\lambda)}$ \\ \hline
$\frac{-2}{s} + \frac{2 (a - 2 b \lambda)}{t_W (s -
m_Z^2)}$&$-\beta_Y$& 0 & $-\beta_Y (1+\frac{s}{2
M_Y^2})$&$-\beta_Y \frac{\sqrt{s}}{M_Y}$\\
$ + \frac{2 (a' - 2 b' \lambda)}{t_W (s -
M_{Z_2}^2)}$ &  &  &  &\\ \hline \hline
\end{tabular}\\[2pt]
where $\beta_Y = (1 - 4 M_Y^2/s)^{\frac{1}{2}}$ and 
$\vartheta$ is the center-of-mass scattering 
 angle between the incident electron momentum $\vec{k}$
and the final $Y^-$ momentum  $\vec{p}$.
 To obtain the amplitude for definite electron helicity 
$\lambda = \pm \frac 1 2$ and definite helicity of
the bilepton ($Y^-$)
$\tau = \pm 1, 0$, the elements in the corresponding
column have to be multiplied by the common factor on top
of the column. 

\end{table}

   We again stress that due to a factor $(1 - 4 s^2_W)$ 
(see Table 1) contribution from the $Z_2$ is very small. 

  Figure 2 shows the dependence of the total cross section
$\sigma (e^+ e^- \rightarrow Y^+ Y^-) $ on the $Y^+$ mass.
We have taken $M_{Z_2} = 1.2 \ {\rm TeV } \  {\rm and}\  3
\ {\rm TeV}$. As we can see from the figure, there is no difference 
between two lines at $M_Y \approx 450$ GeV.

  Now we consider the doubly charged bilepton production:
\begin{equation}
 e^-(k,\lambda) \ + \ e^+(k',\lambda ') \rightarrow 
X^{--}(p,\tau) \ + \ X^{++}(p',\tau ').
\label{pr2}
\end{equation}
The vector currents coupled to $X^{--},
X^{++}$ vanish due to Fermi statistics, therefore 
suitable  Lagrangian for this process is given~\cite{fl}
\begin{equation}
{\cal L} = -\frac{g}{\sqrt{2}} X^{++}_\mu e^T C
\gamma^\mu \gamma_5 e - \frac{g}{\sqrt{2}} X^{--}_\mu
{\bar e} \gamma^\mu \gamma_5 C {\bar e}^T.
\end{equation}
The Feynman diagrams which contribute to this process are shown in Fig.3
\vspace*{0.5cm}

\begin{center}
\begin{picture}(260,50)(-5,0)
\ArrowLine(-10,50)(11,10)
\ArrowLine(11,10)(-10,-30)
\Photon(11,10)(80,10){2}{6}
\Photon(80,10)(100,50){2}{4}
\Photon(80,10)(100,-30){2}{4}
\Text(45,18)[]{$\gamma, Z, Z_2 $}
\Text(-16,54)[]{$e^-$}
\Text(-16,-34)[]{$e^+$}
\Text(110,54)[]{$X^{--}$}
\Text(110,-34)[]{$X^{++}$}
\ArrowLine(150,50)(195,40)
\ArrowLine(195,-20)(150,-30)
\ArrowLine(195,-20)(195,40)
\Photon(195,40)(235,50){2}{4}
\Photon(195,-20)(235,-30){2}{4}
\Text(145,54)[]{$e^- $}
\Text(185,10)[]{$e$}
\Text(145,-34)[]{$e^+$}
\Text(246,54)[]{$X^{--}$}
\Text(246,-34)[]{$X^{++}$}
\Text(90,-80)[]{ Figure 3: Feynman diagram for 
$\ e^+ \ e^- \rightarrow  X^{++} \ X^{--}  $}
\Text(90,-90)[]{ in the  minimal 3 - 3 - 1 model }
\end{picture}
\end{center}
\vspace*{4cm}

  The contribution  from  the electron exchange diagram is given
\begin{equation}
{\cal M}^e_{fi} = -\frac{e^2}{2 t s_W^2}\bar{v}({\bf k}')
\epsilon'\!\!\!/(k\!\!\!/ - p\!\!\!/)\epsilon\!\!\!/
 u({\bf k}).
\end{equation}
Hereafter the notations are similar to those in the previous
case.
\vspace*{0.3cm}

\setcounter{figure}{3}
\begin{figure*}[hbtp]
\begin{center}
\includegraphics[width=10cm,height=8cm]{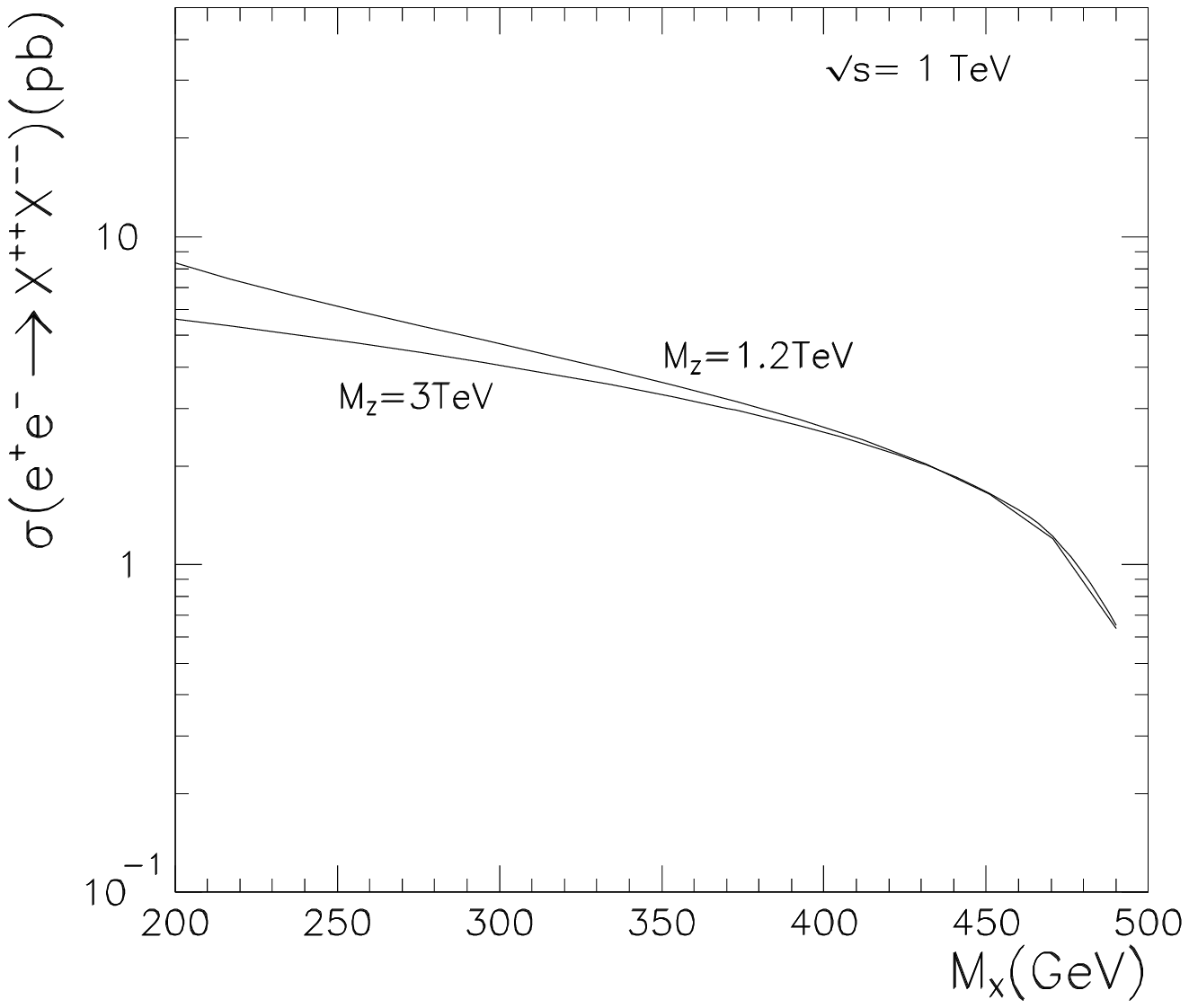}
\caption{\label{lim}{\em Cross section $\sigma (e^+ e^- \rightarrow
X^{++} X^{--}$) in the minimal 3 - 3 - 1 model as a function of $M_X$.}}
\end{center}
\end{figure*}

\begin{table}[htb]
\caption{The helicity amplitudes for
$e^+ \ e^- \rightarrow X^{++} X^{--}$}

\begin{tabular}{|c|c|c|c|c|}  \hline
 &$\tau = \tau' = \pm 1$& $\tau = - \tau' = \pm 1$&
$\tau = \tau' = 0$&$ \tau = 0, \tau' = \pm 1, \epsilon = 1$
 \\&$ -\frac{1}{2} e^2 s \lambda \sin\vartheta$&
$ -\frac{1}{2} e^2 s \lambda \sin\vartheta$&
$ -\frac{1}{2} e^2 s \lambda \sin\vartheta$&
$-\frac{e^2s\lambda}{2\sqrt{2}}(\tau'\cos\vartheta-2\lambda)$
 \\  \hline \hline
$-\frac{1}{2 ts_W^2}$&$ \cos\vartheta -\beta_X$&
$-\cos\vartheta - 2 \tau \lambda$&$  \frac{s}{2 M_X^2}
[\cos\vartheta $&
$ \frac{\sqrt{s}}{2M_X}[\cos\vartheta (1+\beta_X^2)$\\
&    &   & $- \beta_X (1+\frac{2M_X^2}{s})]$    &
$-2\beta_X] -\frac{2M_X}{\sqrt{s}}\frac{\tau'\sin^2\vartheta}{(
\tau'\cos\vartheta -2\lambda)}$ \\ \hline
$\frac{-4}{s} + \frac{2 (a_x - 2 b_x \lambda)}{t_W
(s -m_Z^2)}$& $-\beta_X$& 0 & $-\beta_X (1+\frac{s}
{2M_X^2})$&
$-\beta_X \frac{\sqrt{s}}{M_X}$\\
$+ \frac{2 (a'_x - 2 b'_x \lambda)}{t_W (s -
M_{Z_2}^2)}$ & & & &\\ \hline \hline
\end{tabular}\\[2pt]
\end{table}

The gauge bosons contributions are the same as for the singly charged
bilepton ({\ref{ap1})-(\ref{ap3}) with the couplings $a,b,a',b'$ 
replaced by
\begin{eqnarray}
a_x&=& - \frac{ (- 1 + 4 s_W^2)^2 }{ 4
c_W^2 \sin 2\theta_W},\
b_x =  \frac{( - 1 + 4 s_W^2 ) }{ 4 c_W^2 \sin 2\theta_W},\\
a'_x&=&  \frac{3 (- 1 + 4 s_W^2) }{ 4 c_W^2 \sin 2\theta_W},\
b'_x= \frac{ (1 - 4 s_W^2)}{ 4 c_W^2 \sin 2\theta_W}.
\end{eqnarray}
It is can be verified that in the high energy limit
 $s \gg m_Z^2, M^2_{Z_2}$ the full amplitude vanishes.

The helicity amplitudes for  the considered process
are given in Table 4.

  In figure 4 we plot $\sigma (e^+ e^- \rightarrow X^{++} X^{--}) $ 
as a function of  $M_{X}$ mass.

  Production of the bileptons of the minimal version
at Hadron Collider was considered in~\cite{london}. One found
that the vector bileptons of mass $M_Y \leq 1$ TeV could
be observable at the LHC. Looking at the Table 1 we see that  
contributions from  
the $Z$ and the $Z_2$, due to the factor $(1 - 4 s^2_W) $, 
are very small.  This means that the similar processes
in the model with RH neutrinos will be much bigger.

\subsection{Production  of bilepton in  the  model
with RH neutrinos }

  The amplitude for the singly charged bilepton 
production are obtained from the amplitude 
in the minimal model after modifying coupling constants. 
The  $Z_2 e^+ e^-$ vertex is modified to
\begin{equation}
g^V_{Z_2 ee}  = - \frac{(1 - 4 s_W^2)}{2\sqrt{3 - 4 s_W^2}},\
g^A_{Z_2 ee}  =   \frac{1}{2 \sqrt{3 - 4 s_W^2}}.
\end{equation}
  The amplitudes are simply given in Table 3
replacing $a, b$  by $a_{rhn}, b_{rhn}$ with
\begin{eqnarray}
a_{rhn}&=& \frac{ (- 1 + 4 s_W^2) }{ 4
c_W^2 \tan 2\theta_W},\
b_{rhn} = - \frac{ 1 }{ 4 c_W^2 \tan 2\theta_W},\\
a'_{rhn}&=&  \frac{(- 1 + 4 s_W^2) }{ 4 c_W^2
\sin 2\theta_W},\
b'_{rhn}= \frac{- 1}{ 4 c_W^2 \sin 2\theta_W}.
\end{eqnarray}
Again, it is easy to verify that in the high energy limit
the amplitude for longitudinal bileptons will tend to zero.

\setcounter{figure}{4}
\begin{figure*}[hbtp]
\begin{center}
\includegraphics[width=10cm,height=8cm]{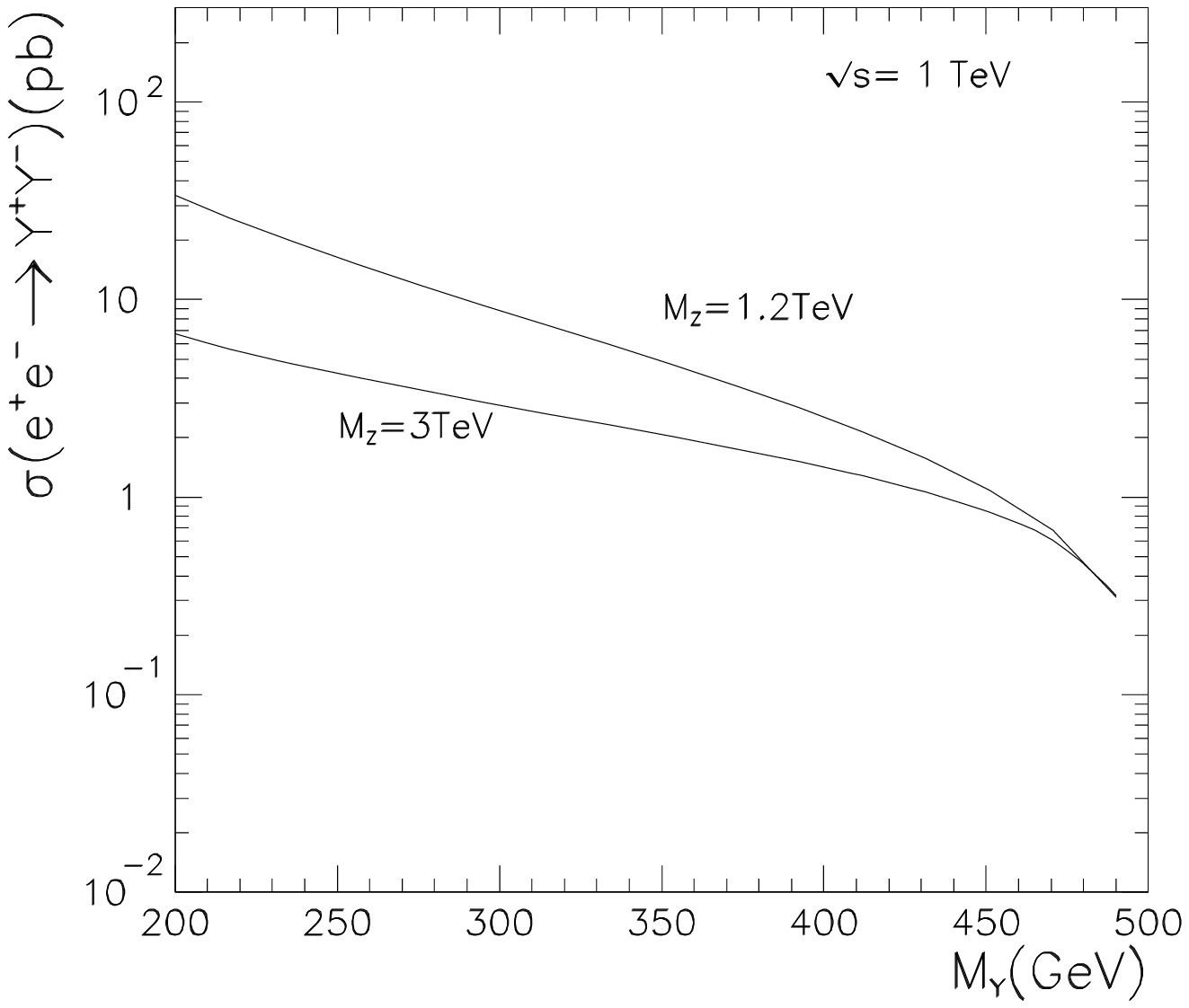}
\caption{\label{lim}{\em Cross section $\sigma (e^+ e^- \rightarrow
Y^+ Y^-$) in the 3 - 3 - 1 model with RH neutrinos as a function of
$M_Y$.}}
\end{center}
\end{figure*}

  Figure 5 shows the dependence of the total cross section
 $\sigma (e^+ e^- \rightarrow Y^+ Y^-) $  in the 3 - 3 - 1 model
with RH neutrinos as a function of $M_Y$.
\vspace*{0.3cm}

\begin{center}
\begin{picture}(260,50)(-5,0)
\ArrowLine(-10,50)(11,10)
\ArrowLine(11,10)(-10,-30)
\Photon(11,10)(80,10){2}{6}
\Photon(80,10)(100,50){2}{4}
\Photon(80,10)(100,-30){2}{4}
\Text(45,18)[]{$ Z, Z_2 $}
\Text(-16,54)[]{$e^-$}
\Text(-16,-34)[]{$e^+$}
\Text(110,54)[]{$X^0$}
\Text(110,-34)[]{$X^{0*}$}
\Text(90,-80)[]{ Figure 6: Feynman diagram for 
$\ e^+ \ e^- \rightarrow  X^0 \ X^{0*}  $}
\Text(90,-90)[]{ in the  3 - 3 - 1 model with RH neutrinos }
\end{picture}
\end{center}

\vspace*{4cm}

  Next, we consider the production of neutral complex
gauge boson in this model
\begin{equation}
e^-(k,\lambda) \ + \ e^+(k',\lambda ') \rightarrow 
X^o(p,\tau) \ + \ X^{o*}(p',\tau ').
\label{pr4}
\end{equation}
For this process we have not only
the photons in the $s$ channel but also neutrino in
$t$ channel (see Feynman diagram depicted in Fig. 6).

 The contributions from $Z$ and $Z_2$ to the amplitude are
similar with those of the previous process after
replacement of  the corresponding mass. Helicity
amplitudes of this process  are given in Table 5.

\begin{table}[htb]
\caption{The helicity amplitudes for
$e^+ \ e^- \rightarrow X^o X^{*o}$}
\begin{tabular}{|c|c|c|c|c|}  \hline
 &$\tau = \tau' = \pm 1$& $\tau = - \tau' = \pm 1$&
$\tau = \tau' = 0$&$ \tau = 0, \tau' = \pm 1,
\epsilon = 1$  \\
 &$ -\frac{1}{2} e^2 s \lambda \sin\vartheta$&
$ -\frac{1}{2} e^2 s \lambda \sin\vartheta$&
$ -\frac{1}{2} e^2 s \lambda \sin\vartheta$&
$-\frac{e^2s\lambda}{2\sqrt{2}}
(\tau'\cos\vartheta-2\lambda)$ \\  \hline \hline
$ \frac{2 (a_{xo} - 2 b_{xo}
\lambda)}{t_W (s - m_Z^2)}$&
$-\beta_{Xo}$& 0 & $-\beta_{Xo}
(1+\frac{s}{2 M_{Xo}^2})$&
$-\beta_{Xo} \frac{\sqrt{s}}{M_{Xo}}$\\
$ + \frac{2 (a'_{xo} - 2 b'_{xo} \lambda)}{t_W (s -
M_{Z_2}^2)}$  & & & & \\ \hline \hline
\end{tabular}\\[2pt]
where
\begin{eqnarray}
a_{xo}&=& \frac{ ( 1 - 4 s_W^2) }{ 4
c_W^2 \sin 2\theta_W},\
b_{xo} =  \frac{ 1 }{ 4 c_W^2 \sin 2\theta_W},\\
a'_{xo}&=&  \frac{(- 1 + 4 s_W^2) }{ 4 c_W^2
\sin 2\theta_W},\
b'_{xo}= \frac{- 1}{ 4 c_W^2 \sin 2\theta_W}.
\end{eqnarray}
\end{table}

Applying formula (120) in~\cite{cd} we get the cross section
for this process
\be
\sigma(e^+ e^- \rightarrow X^o X^{o*}) = \frac{\pi
\alpha^2}{ 6 s}\beta_X^2\left( \frac{4}{1-\beta_X^2} -
1 - 3\beta_X^2 \right)\Sigma_{X^o},
\label{sig1}
\ee
where
\begin{eqnarray}
\Sigma_{X^o} & = & | L L|\frac{s^2(1-2s_W^2)^2}{4
\sin^2  2\theta_W}
 \left[ \frac{1}{(s - m_Z^2)} -
\frac{2}{(s - M_{Z_2}^2)}\right]^2 \nonumber\\
 &  &+| R R|\frac{ s^2 t_W^2}{4} \left[
\frac{1}{(s - m_Z^2)} -
\frac{1}{(s - M_{Z_2}^2)}\right]^2,
\end{eqnarray}
and
\begin{eqnarray}
| R R| & = & \frac{1 + P_+ + P_- + P_+ P_-}{4},\nonumber\\
| L L| & = & \frac{1 - P_+ - P_- + P_+ P_-}{4}.
\end{eqnarray}

In the  limit $|P_{e^-}| = |P_{e^+}| = 1, s_W^2 = 0.25$,
Eq. (\ref{sig1}) becomes:
\begin{eqnarray}
\sigma(e_R^+ e_R^- \rightarrow X^o X^{o*})
&=& \frac{\pi \alpha^2\ s}{24}\beta_X^2 t_W^2 \left(
\frac{4}{1-\beta_X^2} - 1 - 3\beta_X^2 \right)\nonumber\\
& &\times \left[ \frac{1}{(s - m_Z^2)} -
\frac{1}{(s - M_{Z_2}^2)}\right]^2.
\end{eqnarray}

\setcounter{figure}{6}
\begin{figure*}[hbtp]
\begin{center}
\includegraphics[width=10cm,height=8cm]{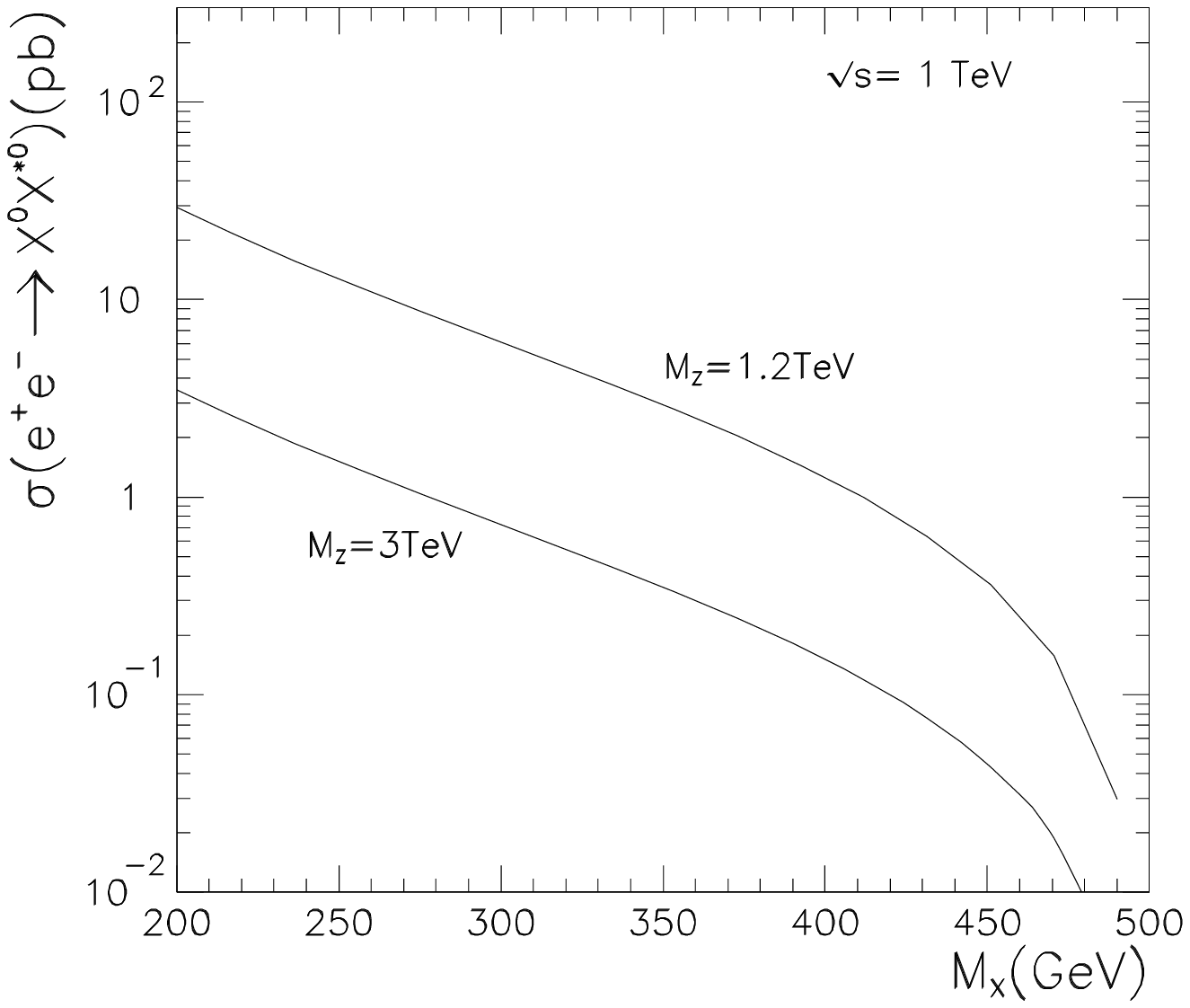}
\caption{\label{lim}{\em Cross section $\sigma (e^+ e^- \rightarrow
X^0 X^{0*}$) in the 3 - 3 - 1 model with RH neutrinos as a
 function of $M_X$.}}
\end{center}
\end{figure*}

 The cross section for this process is plotted in Fig. 7.
From Table 2 we see that the contribution from $Z_2$ is much bigger
than that in the minimal version. As we can see from  figures
2, 4, 5 and 7, when $M_{Z_2}$ is not too heavy
the cross sections in the model with RH neutrinos
can be one order bigger than that in the minimal version.

  We ignored the questions connected with the experimental
difficulties of identifying the neutral gauge bosons, which 
interact with neutrinos (and the exotic quarks) only.

\section{Discussion and numerical results }

  From the helicity amplitudes in the previous section it is a 
simple task to compute the cross section for production
of any pair of bileptons in a given helicity state.
In figures 2 and 4, we show the total cross section for
production of X and Y in the minimal model at 1~TeV.
With planned colliders of luminosity ${\cal L}=80 fb^{-1}$,
one expects several thousand events almost up to the
kinematic limit. 
Note that the $Z_2$ coupling to $Y$ is proportional to $1-4\sin^2\theta$
so the pair production process is not very sensitive to the 
$Z_2$ exchange.

  Doubly charged bileptons can also be produced singly
in $e^+e^-$ via Weisaker-Williams photons and the subprocess
$e\gamma\rightarrow Xe$~\cite{rizzo}.
In this model  where the bilepton coupling is of electromagnetic
strength, the pair
production is slightly larger
 in the mass range kinematically allowed.
We note here that the coupling of the single charged bileptons are
exactly the same as for the W except for the third generation
where the $Y$ couples to the b quark and an exotic up-type quark.
Therefore if the exotic quark is heavier than the bileptons
we have the decay width
$\Gamma_Y = \frac{G_F m^2_W M_Y}{2\sqrt{2} \pi}$.

  The total cross section for new gauge bosons pair 
production in the right-handed model are given
in Fig.~5.
The production cross section is sizeable, and
at least for singly charged bileptons, a signal 
should easily be extracted from either the purely
leptonic decay mode 
or the semileptonic mode.
For the complex neutral bileptons, extracting a signal over
the background is more troublesome as 
the leptonic decay mode is exclusively into neutrinos.
As for the $Y$, the neutral bileptons couples to the
first two generations of quarks and to a pair
of top quark/exotic quark with equal strength.

  Since the $Z_2$ turn out to be heavier than two bileptons
it should be  seen four jet final states that is unique
signature of the considered minimal model. 
As expected, the linear collider sensitivity to $Z_2$ properties 
is best when running near the resonance $\sqrt{s}= M_{Z_2}$.
With the  planned machine parameters the considered processes 
should be seen unless the bilepton masses are bigger than 2 TeV.

  Finally, it must be emphasized again that: 
when $M_{Z_2}$ is not too heavy the cross sections
in the version with RH neutrinos can be one order bigger than
the same in the minimal model.
\vspace*{0.2cm}

{\bf Acknowledgments }

  The authors express their sincere 
thanks to  G. Belanger for stimulating conversations 
and help in numerical calculation.  
They also thank J. L. Kneur
for providing the calculation program.
One of the authors (H. N. L) would like to thank P.
Aurenche, F. Boudjema, J. Kaplan for helpful discussions.
He would also like to thank Theory Group, LAPP for warm 
hospitality and CNRS for financial support under the joint
France-Vietnam Convention Internationale.
 This work is supported in 
part by the Natural Science Council of Vietnam.
\\[0.5cm]

\end{document}